# DOES CONSCIOUSNESS COLLAPSE THE WAVE FUNCTION

Dick J. Bierman
University of Amsterdam

**Abstract**

A conceptual replication of the Hall-experiment to test the 'subjective reduction' interpretation of the measurement problem in Quantum Physics is reported. Two improvements are introduced. First the delay between pre-observation and final observation of the same quantum event is increased from a few microseconds in the original experiment to 1 second in this replication. Second, rather than using the observers conscious response as the dependent variable, we use the early brain responses as measured by EEG. These early responses cover a period where the observer is not yet conscious of the quantum event. Results support the 'subjective reduction' hypothesis because significant difference between the brain responses of the final observer are found dependent upon the pre-observer looking or not looking at the quantum event (exact binomial p < 0.02). Alternative 'normal' explanations are discussed and rejected. It is concluded that the present results do justify further research along these lines.

## 1 Introduction

1.1 The Problem

In 1977 Hall et al (Hall et al, 1977)) reported an experiment that, according to their description, tested the most radical solution to the 'measurement problem' in quantum physics, namely the proposition that : …. *The reduction of the wave packet is a physical event which occurs only when there is an interaction between the physical measuring apparatus and the psyche of some observer…..*

They defended their experiment writing: … *although we concur that there is a genuine problem of the reduction of the wave packet, we do not intend in our paper to defend this opinion against those who maintain that it is a pseudo problem…..*

In spite of many attempts, like the relative state solution (Everett, 1957) and the introduction of non-linear terms in the Schrödinger equation (Ghiradi, 1986), the measurement problem seems still not be solved. This failure to clearly resolve the problem has left the physics community polarized with some contending the problem remains a fundamental shortcoming in the quantum formalism and others holding that there is no reduction of the wave packet at all (Bohm and Hiley, 1997; Griffith, 2002; Dieks and Vermaas, 1998). Costa de Beauregard (1976), Walker (1971, 1988, 2000) and later Stapp (1993) have argued, using arguments provided by a.o. von Neumann (1955) and Wigner (1967), that none of these solutions are acceptable and that subjective reduction is still a possible and even preferred alternative.

We, like Hall et al, do not wish to fight this or any other position with regard to the proper interpretation of the quantum formalism and the role of the measurement therein, but like Hall and his collaborators, we would like to investigate this problem experimentally.

1.2 Hall's experiment

The Hall experiment is conceptually easy to understand. A quantum event, in this case a radioactive decay, is measured in a counter and the signal is displayed on a scaler. An observer 1 is observing the scaler. The scaler signal is transmitted through a delay unit and displayed again. The second scaler is observed by 'final' observer 2.

Observer 1 will **sometimes** observe but sometimes not observe his scaler. Observer 2 had to 'guess' if a quantum event observed by him has already been observed by observer 1 (see fig.1 from the original publication)

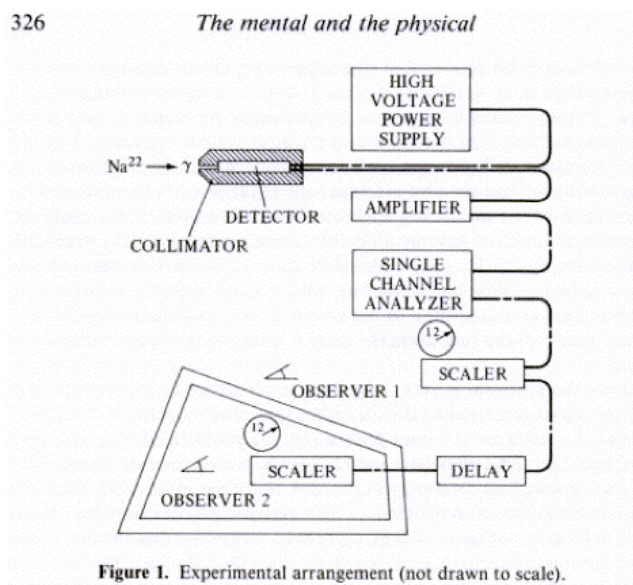

Figure 1. Experimental arrangement (not drawn to scale).

The results of this experiment were precisely at chance. I.e. the second observer guessed 50% correct. Hence it was obvious that this observer was unable to detect if the observed signal had already been observed earlier. It was concluded that the experiment does not provide support for the hypothesis that it is the interaction with consciousness that causes the wave packet to collapse.

1.3 The present conceptual replication

It should be made explicit that there is an implicit assumption of Hall et al that our brains in some way are able to detect the difference between a superposition state and a singular state. And also that this difference can be communicated consciously.

In a comment later added to the article the authors note that the used delay was extremely short and that …. *The delay time should be in the order of psychologically discriminable intervals……*Since the original Hall paper has been widely acknowledged as evidence against the hypothesis that it is the interaction with consciousness that causes the wave packet to collapse, it is essential that this serious error in the Hall experiment be corrected by means of further testing. It is important that both the time delay be of a physiological significant duration and that the determination as to whether the observers have been affected differently by the two conditions be placed on a more objective foundation than the verbal report used by Hall.

In the present conceptual replication the time between the first observation and the second one was set to 1000 msecs. Indeed Libet's seminal work on the processing time needed for conscious experience sets a lower interval of about 300-500 msecs because one should require the first observation to be a conscious one (Libet, 1991). The difference with the original experiment goes a bit further than just adjusting this interval.

Rather than asking the second observer for a conscious guess we measure the brain responses to the stimulus. This is done for the following reason: If consciousness is the crucial element for wave packet reduction, the conscious decision, used as dependent variable in the original experiment, will be based on the physical state of the wave packet *after* consciousness in the second observer has developed. At that time, the wave packet according to the hypothesis under investigation, has already collapsed even if no pre-observation has taken place. Thus, the manipulation of the pre-observer will not induce any difference in the final observer with regard to his conscious behavior.

By measuring brain potentials of the second observer one can however also tap into the early (< 300 msec) non conscious processing of the brain. At that time the wave packet is supposedly still in superposition, but only if no pre-observation has taken place.

## 2 Design of the experiment

Quantum events were generated by an alpha particle source (as used in smoke detectors; 2P40-76-18) that was mounted on slider allowing the source to be moved with respect to a lead shielded Geiger-Muller counter (Automess 6150-100). The distance was set so that on the average 1 particle about every second was detected. The counter pulse was amplified and fed to the trigger channel of an EEG data-acquisition system (*Biosemi Active-1*, 2003). We used *National Instruments LabView* software (NI, 2003) to detect this trigger and to transform it into a delayed audio beep of 1500 Hz and 50 msecs duration. The audio-delay was 1 second. The software randomly would generate a visual stimulus of ~65 msecs duration directly upon the trigger. The visual stimulus therefore preceeds the audio-beep by a time sufficient for the first observer for conscious experience of the quantum event *before* the second observer (see fig.2). The random decision to show this visual stimulus to the first observer before submitting the beep to the second observer or not was pseudo random with the seed determined by the computer clock. After each quantum event thus measured there was a dead time of 2 seconds during which the input of the Geiger-Muller counter was discarded. The subjects were asked to count the number of observed quantum events.

The quantum mechanical theory of radioactive decay describes the emitting particle as a superposition of two states, the decayed and the non-decayed state. Although our measurement system is a composite many particle system it can be regarded as being in a superposition of a 'decayed' and a 'non-decayed' state. According to the radical proposition under consideration, a reduction of this superposition occurs only when an observer 'looks' at one of the two indicators of the emission. Either observation of the visual or the audio representation would collapse the wave packet.

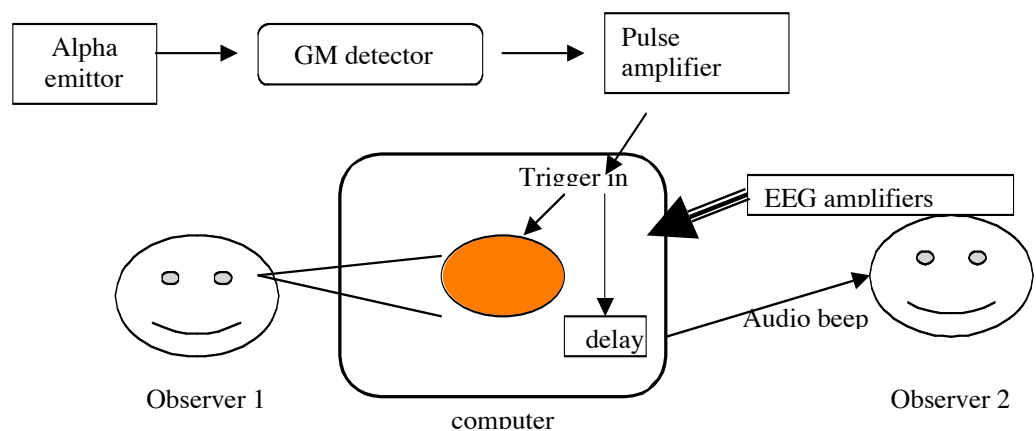

Figure 2. The experimental set-up of the present replication experiment. (Note that this figure doesn't show the video connection from Observer 2 to Observer1).

## 3 Experimental Procedure

*Subjects*
Volunteer subjects were invited in pairs. These were generally freshman psychology students who participated for course credit. In total 9 male and 21 females, providing useful data, participated in the experiment (mean age=21.4, sd=4.7). Two subjects were removed from the analysis because the brain signals they provided were noisy due to a loose electrode.
   Upon arrival they were fully informed about the purpose and potential implication of the experiment. First they participated in a so-called odd-ball task used to test the equipment. Then the crucial task, which we, for obvious reasons, called the '*Schrödinger-task*' was presented. The role of observer 1 and 2 were played by both subjects in two separate runs.

*Physiological measurement and further procedure.*
Sintered AgCl EEG electrodes with active preamplifiers (*Biosemi Active 1*) were connected to the head of observer-2 using the standardized 10/20 system (see appendix 1). No temporal electrodes were used. Then observer 2 went to a neighboring room and was seated in a relaxing chair while observer 1 stayed at the computer screen with the experimenter. A short 'calibration' experiment was run consisting of the above mentioned odd-ball task in which observer-2 was presented each 3 seconds (with one second random jitter) audio beeps of 30 msecs duration. Hundred beeps with either a frequency of 1200 Hz or a frequency of 2000 Hz were presented. The frequency was randomly determined with the probability for the higher frequency 4 times as low as for the lower. The subject was asked to count the higher frequency beeps. Their brain signals were sampled with a frequency of 2400 samples per second. If the resulting average evoked brain potentials conformed to the well know average auditory brain potential (Picton et al, 1974), the actual 'Schrödinger' run was started with observer 1 sitting in front of the computer screen observing the visual stimulus that appeared in about 50% of the cases directly upon a radioactive decay. The experimenter refrained from looking at this screen. The total run consisted of 120 radio-active decay events. This took about 8 minutes.

After a short break roles were switched and the procedure was repeated. The total experiment took less than one hour.

To prevent ourselves from data snooping and data selection with the goal to 'find what we were searching for', we first analyzed the results of the standard, and completely unrelated, odd-ball task. Once we had fixed the complete procedure on the basis of exploration of these odd-ball task data we would allow ourselves to analyze the actual data.

On the basis of the explorations of the odd-ball data it was concluded that two of the 32 subjects were not providing valid EEG data. Furthermore these odd-ball data were used to establish an optimal preprocessing procedure. The thus established preprocessing procedure consisted of 4 steps. All signals were re-referenced to (compared to) the signal at the Pz electrode. First a 50 Hz notch filter was applied, then the data were filtered through a band pass filter between 1 and 30 Hz (slopes = 24 db/Oct). Then (eye) movement artifacts were removed from the data. The criteria used were 'absolute value' and 'derivative'. On the average this algorithm removed about 5-10% of the available segments. Because there is a high correlation between the results obtained from different leads (electrodes) we did a factor analysis to see how we could combine the signals of different leads into a compound signal. This analysis gave two clear factors, one consisting of the central and frontal leads with a mean factor load of 0.87 and one for the parietal electrodes with a mean factor load of 0.93. We called the two combined signals 'FC' which is the average of 11 leads (Fpz, Fp1, Fp2, Fz, F7, F3, F8, F4, Cz, C3, C4) and 'P' which is the average of two leads (P3 and P4).

## 4 Results

After having thus established the data to be used, and a well specified preprocessing procedure we applied this without further adaptation to the EEG data obtained in the 'Schrödinger' part of the experiment. First we averaged all the data pooled over subjects and pre-observer condition per sample for the FC- and P-signal separately.

The thus obtained evoked potentials were submitted to the automatic peak detection procedure of the standard BrainVision analyzer software. Results are given in fig. 3.

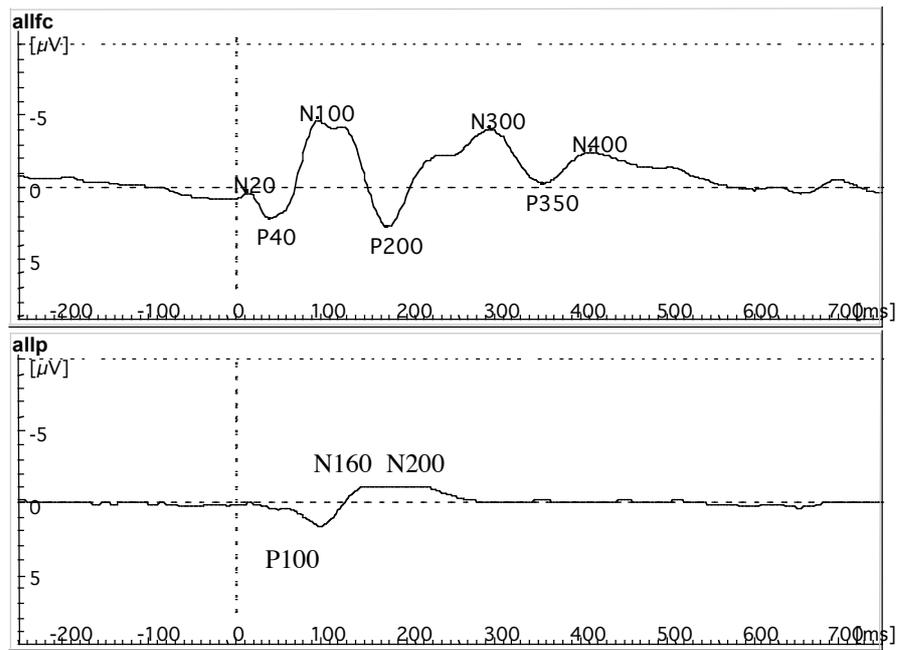

Figure 3. The mean evoked response for all subjects. Upper trace: all frontal and central leads (FC leads). Lower trace: two parietal leads (P-leads).

Note that in figure 3 the evoked potentials are still pooled over pre-observer conditions. In order to check if the pre-observation by a first observer makes a difference for the brain signals of the final observer we have to split the data for the two pre-observer conditions.

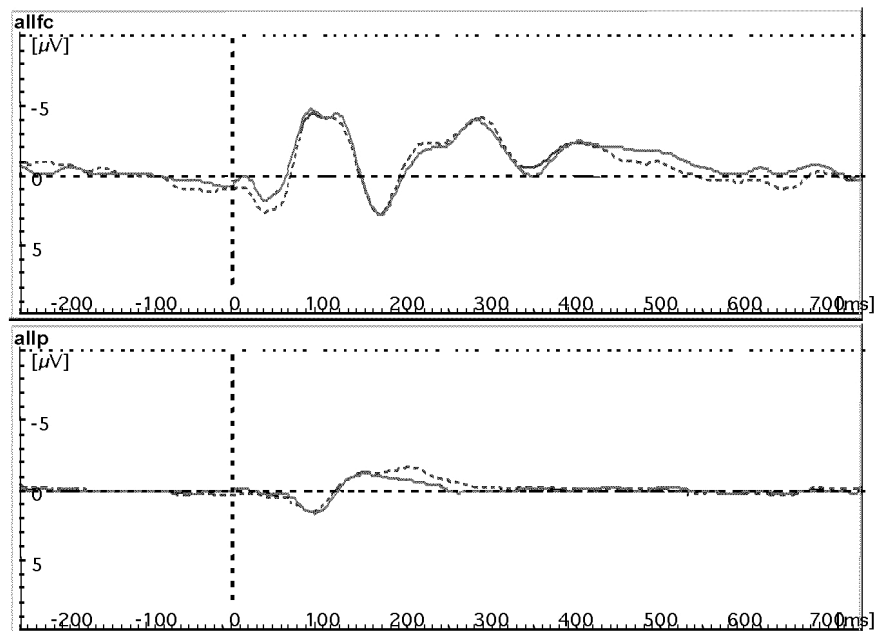

Figure 4. The mean evoked responses split for pre-observation condition
(solid = pre-observed; - - - = not pre-observed).

In figure 4 the same evoked potentials are plotted but this time separately for the two pre-observer conditions. Under the null-hypotheses (that pre-observation doesn't matter) the difference between the two curves should be nill.

**Statistical Analysis of peak amplitudes**

As usual in these EEG data, the two traces for the two conditions do not completely coincide. In order to assess if the observed differences are statistically meaningful we did a simple comparison between the signal value at peak position for the pre-observed and the non pre-observed trials.

All peaks obtained by the automatic peak detection procedure were analyzed:

For the combined frontal and central leads: N20, P40, N100, P200, N300, P350 and N400. At exact 17, 41, 95, 178, 292, 357 and 411 msecs after stimulus onset. (The convention in EEG plots is generally that positive voltage is plotted "down", i.e. to the bottom of the page.) For the two combined parietal leads, P100, N170 and N200 at exact 99, 160 and 212 msecs after stimulus onset.

In Table I, column 3, we give the differences for the peak amplitudes between the two observer-conditions. As said before these differences should be negligible under the assumption that the fact that someone has observed the same quantum event earlier doesn't matter. A standard t-test was run to find the probabilities that the observed differences are due to chance (column 5). In order to assess the over-all probability of these differences between the peaks we calculated the exact binomial probability of finding three significant differences in a total of 10 analyses. The over-all probability thus calculated is p= 0.0115.

Besides of the results of the parametric t-test we also calculated the results of the non-parametric binomial tests. In this latter test the magnitude of the difference is not relevant, only the direction for each subject. It can be argued that the non-parametric test is more suitable since the differences between two evoked potentials are not necessarily normally distributed. An overall assessment of the non parametric tests using the same binomial approach as above gives a total p-value of p= 0.0106.

|  | Peak | Difference (microvolts) | df = 29 t | p | Non-parm p N=30 |
|---|---|---|---|---|---|
| FC-leads | N20 | 1.002 | 2.12 | *0.043* | 19-11: 0.20 |
|  | P40 | 0.903 | 2.64 | *0.013* | 22-8: 0.016 |
|  | N100 | 0.350 | 0.66 | 0.52 | 15-15 |
|  | P200 | -0.09 | -0.18 | 0.86 | 15-15 |
|  | N300 | -0.04 | -0.08 | 0.93 | 15-15 |
|  | P350 | -0.54 | -1.17 | 0.25 | 12-18: 0.36 |
|  | N400 | 0.098 | 0.25 | 0.80 | 16-14: 0.86 |
| P-leads | P100 | -0.16 | -0.67 | 0.50 | 12-18: 0.36 |
|  | N160 | -0.152 | -0.84 | 0.41 | 13-17: 0.58 |
|  | N200 | -0.956 | -3.93 | *0.0005* | 7-23: 0.005 |

Table I: Results of the differential analysis of the peak amplitudes.

From these results the following conclusions may be drawn:

1. With regard to the signal from frontal and central leads there is a significant difference between the conditions in the very early peaks. This difference is gone after about 100 milliseconds.

2. On the parietal leads the difference is into the other direction and arises later with a clear maximum at 200 milliseconds.

# 5 Discussion

The results of these experiments do support a solution of the measurement problem that gives a special status for conscious observation in the measurement process. The absence of significant differences in the late evoked potential appears to be in line with the fact that in the original Hall experiment no differences were found when one asked the second observer to *consciously* express his feeling if the observed quantum event had already been observed. This finding however should be treated cautiously because the lack of statistical power in the later phases of the response. This lack of power is caused by increased variance with increasing latency times.

Before drawing far reaching conclusions we should first check if there are no more mundane explanations for the current findings.

*Alternative explanations*

Spurious sensory cueing of the second observer has been considered. The reason for having the first observer observe a *visual* representation of the quantum event rather than a audio-beep was indeed to prevent any audio leakage to the second observer. Both observers were in adjacent and not auditory or electromagnetically shielded rooms. Ultrasonic or electromagnetic signatures from the monitor displaying the signal for the first observer might still have presented sensory cues. Thus the second observer might have produced a slightly different auditory evoked potential due to this earlier pre-observation related ultra-sound. This scenario, however, is not very plausible in that it would result in affecting the peaks in the evoked potential in a systematic way. The timing of the visual stimulus to the first observer and the delayed audio beep to the second is not precise and therefore one can hardly expect a well-defined effect in time.

A second explanation might be found in improper randomization of the pre-observer condition. It is well known that evoked potentials on simple stimuli like beeps tend to habituate (decline). Thus the amplitude of the signal becomes smaller in the course of the experiment. If, for some reason, the randomization did result in a non-balanced distribution of conditions in time this could artificially induce a differential effect due to habituation. We tested this idea using the actual sequence of stimulus conditions as they occurred in the experiment with several habituation models. None of these models gave any effect (p-values around 0.77). As a further test on the validity of the peak differences that we found between the two pre-observer conditions, we 'randomly' relabeled the markers so that we created two pseudo-conditions for which we did exactly the same peak difference analysis. The result of this analysis was at chance level. (the mean difference found was 0.16 microvolts at the P40. This is 6 times smaller than the real effect).

Although the current results look pretty robust, they are not *extremely* improbable in terms of statistics. Although the applied over-all assessment is a bit conservative since it treats the more significant peaks as being of the same significance as the least significant peak, one could argue that the current findings might be attributable to chance with a probability of 1 in 50. Although this figure satisfies the criterion of 5% which is generally accepted as the significance criterion, it is not enough to unequivocally accept the hypothesis that consciousness collapses the state vector. Strong claims need strong evidence.

*Further work*

The further crucial experiment in which the radioactive source is replaced by a pseudorandom source is presently underway. In this experiment, the differential effect should disappear in this latter condition as the quantum character of the observed event is crucial. This further work will be reported subsequently.

In these replication studies we now are also able to predict more precisely where and when to look for differences in the brain signals.

The role of the video-camera which was brought in to ensure that an interaction of 'states' of both observers was entering the state description of the experiment is another factor that can be explored. If such a camera is necessary, it would follow that the current set-up could not be used to 'send signals' outside of the light cone and hence does not violate relativity theory.

So far the concept of a conscious observation has not been worked out in detail. In Libet's work, which we used to estimate the delay between perceptual input and the conscious experience thereof, the conscious observation is by definition an observation which is stored in memory. However there is suggestive evidence, for instance from 'change blindness' experiments, that there is another form of 'faster' conscious experience directly related to perceptual input (Landman et al, 2003). This experience is not stored in memory. In further work it might be necessary to discriminate between these and possibly other forms of conscious experience.

In work in the field of 'Artificial Intelligence', the question has arisen if future computers might become conscious. The present results suggest that such a question can become empirically testable.

*Acknowledgements*

*BioSemi* offered generous support by loaning the EEG equipment. Chris Duif was helpful in setting up the software. Ronald van der Ham ran this experiment as a part of his master's thesis. Dennis Dieks helped to understand and describe the experiment in a formal way. The experiment was originally designed at *Starlab*. All former Starlab personnel are thanked for providing the unique climate for real scientific enquiry.

*Raw Data:*
http://a1162.fmg.uva.nl/~djb/research/eeg_data

*Equipment:*
National Instruments' Labview (2003) http://www.ni.com

The 10-20 electrode placement system:
http://faculty.washington.edu/chudler/1020.html

Biosemi Active-1, (2003):
http://www.biosemi.com/

---

APPENDIX I
The 10/20 EEG electrode placement system

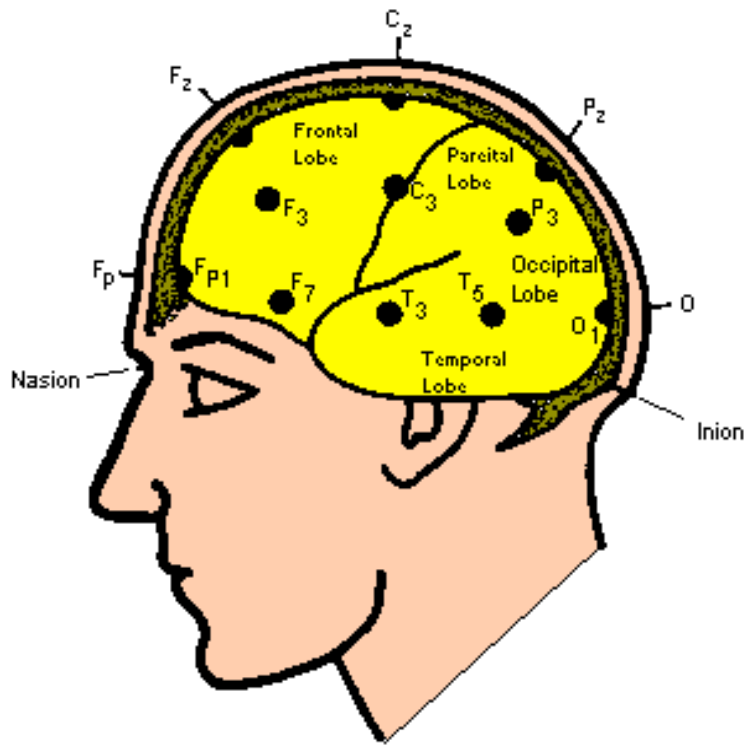